# Negative ion density in the ion source SPIDER in Cs free conditions

M. Barbisan,[a,1] R. Agnello[b], G. Casati[c], R. Pasqualotto[a], C. Poggi[a], E. Sartori[a], M. Spolaore[a] and G. Serianni[a]

[a] *Consorzio RFX, corso Stati Uniti 4 – 35127 Padova, Italy*

[b] *École Polytechnique Fédérale de Lausanne (EPFL), Swiss Plasma Center (SPC), CH-1015 Lausanne, Switzerland*

[c] *Imperial College London, Exhibition Rd., South Kensington, SW7 2BX, London UK*

E-mail: marco.barbisan@igi.cnr.it

ABSTRACT: The SPIDER experiment, operated at the Neutral Beam Test Facility of Consorzio RFX, Padua, hosts the prototype of the $H^-/D^-$ ion source for the ITER neutral beam injectors. The maximization of the ion current extracted from the source and the minimization of the amount of co-extracted electrons are among the most relevant targets to accomplish. The Cavity Ring-Down Spectroscopy diagnostic measures the negative ion density in the source close to the plasma grid (the plasma-facing grid of the ion acceleration system), so to identify the source operational parameters that maximize the amount of negative ions which can be extracted. In this study SPIDER was operated in hydrogen and deuterium in Cs-free conditions, therefore negative ions were mostly produced by reactions in the plasma volume. This work shows how the magnetic filter field and the bias currents, present in SPIDER to limit the amount of co-extracted electrons, affect the density of negative ions available for extraction. The results indicate that the magnetic filter field in front of the acceleration system should be set between about 1.6 mT, condition that maximizes the density of available negative ions, and about 3.2 mT, condition that minimizes the ratio of electron current to ion current. The negative ion density also resulted to be maximized when the plasma grid and its surrounding bias plate was positively biased against the source body with a total current in the range 0 A÷100 A. The paper shows also how much, in Cs-free conditions, the electric fields in the acceleration system can affect the density of negative ions in the source, close to the plasma grid apertures.

# 1. Introduction

ITER Heating Neutral Beam injectors (HNBs) are based on sources of negative hydrogen or deuterium ions and will provide 16.7 MW beams of particles at about 1 MeV energy. The present design of these ion sources is based on a radiofrequency (RF) inductively coupled plasma (ICP) generation concept [[1]-[4]]. The negative ions are extracted from the plasma box and accelerated by a system of grids. The full-scale prototype of ITER HNB ion sources is currently in operation in the SPIDER experiment, which is part of the Neutral Beam Test Facility (NBTF) located at Consorzio RFX (Padua, Italy). The ITER HNB ion sources and their prototype, together with their respective acceleration system, are required to deliver a beam current density of 355 A/m$^2$ (H)/285 A/m$^2$ (D) from a total extraction area of 0.2 m$^2$, and with a ratio between co-extracted electrons and negative ions lower than 0.5 (H)/1 (D), for a beam duration of 1 h [[1]-[4]]. Negative ions can be produced in the plasma volume by dissociative electron attachment reactions involving rovibrationally excited hydrogen or deuterium molecules. A much higher amount of negative ions can be made available if caesium (Cs) is evaporated into the source: the surface work function is lowered, so that the surface processes that convert H/D and positive ions into H$^-$/D$^-$ exponentially increase, becoming dominant in negative ion production [[5]-[8]]. Because of the quasi-neutrality of the plasma, the increase of negative ion density close to these apertures is also beneficial in limiting the local density of electrons, which could then be co-extracted. The technological challenges and the need to understand and optimize the plasma properties in such a large source required to operate SPIDER first in Cs-free conditions, from May 2018 to April 2021. In absence of Cs the negative ion density at the accelerator apertures is not sufficient to reach the required absolute values of extracted beam current density; it is however possible to find the best source operation conditions to maximize negative ion production, as starting point for the subsequent experimental campaigns with Cs evaporation. Measuring the negative ion density is possible by means of a Cavity Ring-Down Spectroscopy [9] (CRDS) diagnostic, installed in SPIDER in 2020 [[10],[11]]. The SPIDER CRDS employs 1064 nm laser pulses that are trapped in an optical cavity, composed by two high reflectivity (>99.99%) mirrors; the cavity axis crosses the region from which negative ions are extracted from the plasma. Some photons γ are absorbed by the negative ions in photodetachment reactions (H$^-$/D$^-$+γ→H/D+e); thanks to the high reflectivity of the mirrors, the absorption path length is multiplied thousands of times. By measuring the intensity decay of the light pulses exiting the cavity it is possible to estimate the negative ion density. This technique is used in several negative ion sources for fusion [[12]-[20]]. The CRDS diagnostic setup, the data analysis algorithm and the first experimental results were presented in [11]. Sections 2 of this paper will recall the structure of the SPIDER source, the physical principles at the base of its functioning and the parameters used to control it. Section 3 will recall the structure and the data analysis technique of the SPIDER CRDS diagnostic. At last, sec. 4 will present and discuss the latest CRDS measurements in Cs-free conditions, both in hydrogen and deuterium.

## 2. The SPIDER ion source

A lateral view of the SPIDER negative ion source is schematized in Figure 1a. Following a modular concept, eight cylindrical ICP plasma sources (called plasma drivers) are present, arranged in four rows and two columns; each row of sources is powered by a separate RF generator (max. 200 kW at 1 MHz), labelled RF1-RF4 [21]. All the drivers let the plasma diffuse in a common expansion chamber, towards the acceleration system; a system of permanent magnets creates a multicusp filter field at the source walls, reducing plasma losses. The need to minimize the negative ion losses inside the acceleration system of the ITER HNB, due to collisions with gas molecules, poses an upper limit on the gas pressure that can be set inside the source. As a compromise with the beam performance requirements mentioned in sec. 1, the experimentation on SPIDER is aimed at a target source pressure of 0.3 Pa [2].

Similarly to what experimentally found in analogous experiments, the plasma in the drivers and in the upstream part of the expansion region are expected to reach an electron density $n_e$ and an electron temperature $T_e$ up to about $10^{18}$ m$^{-3}$ and 10 eV, respectively [[23]-[25]]. In Cs-free operation, this plasma region contributes to provide ro-vibrationally excited $H_2/D_2$ molecules, which can then undergo dissociative attachment reactions in the low temperature region of the plasma source, leading to the generation of negative ions. With Cs evaporation, the driver region will be essential to provide H/D neutrals and positive ions, to be converted into negative ions at the surface of the Plasma-facing Grid (PG), the first electrode of the acceleration system [6]. In the expansion region, a rather uniform transverse magnetic field is generated by a current $I_{PG}$, which flows vertically in the PG (1.6 mT close to the PG per 1 kA of current, up to 8 mT) [22]; $I_{PG}$ normally flows downwards, but its direction has been occasionally reversed. This magnetic filter field causes the electrons to be magnetized, thus reducing the plasma diffusion towards the PG especially for hot electrons, obtaining rather different plasma properties in the proximity of the PG: a density of few $10^{17}$ m$^{-3}$ and an electron temperature below 2 eV, respectively, as measured in a similar ion source [[6],[23],[25]-[27]]. In SPIDER, these quantities are measured by RF compensated Langmuir probes, installed in the PG region [[28]-[31]]; the measurements are in agreement with different typologies of probes, installed for a limited time in SPIDER [32], and also in basic agreement with what measured by electrostatic probes in other RF sources [[25],[33]]. The reduction of electron temperature prevents the destruction of negative ions by electron stripping [34]. The reduction of $n_e$ and $T_e$ also diminishes the amount of electrons that could get sufficiently closer to the PG apertures to be co-extracted with the negative ions [35].

To further limit the density of electrons in the plasma at the extraction apertures, the PG is positively biased against the source body, to attract and remove electrons at the PG; the PG bias can reduce or even reverse the potential drop at the PG plasma sheath [[6],[12],[35]-[39]]. For the same purpose the bias plate, an electrode located 10 mm upstream the PG and framing the groups of PG apertures (represented in Figure 1b in light blue), can be independently and positively biased against the source body. The biasing of PG and BP is respectively performed by the ISBI (Ion Source Bias) and ISBP (Ion Source Bias Plate) power supplies, which are normally current-regulated in order to control the net flux of removed electric charges. Both ISBI and ISBP have a resistor R=0.63 Ω in parallel to better handle the transient conditions imposed by the plasma, that may cause damage to the power supplies (eg. polarization reversal). All the bias currents presented in this paper represent the actual currents flowing between source body and PG or between source body and BP, i.e. the ISBI and ISBP currents minus the currents flowing in the respective resistors in parallel.

The use of a magnetic filter field in a negative ion source was demonstrated to cause cross-B drifts of the plasma in the expansion region [33], phenomenon that can be accentuated by the electric fields generated in the BP-PG region by the electric biasing of these elements. A comprehensive accounting and understanding of the effects of the magnetic filter field on the transport of plasma particles can only be reached by means of numerical simulations, as shown for example in refs. [[40]-[43]]. In a similar RF negative ion source it was observed that by inverting the magnetic field direction and thus the drift direction, the non-uniformities on the plasma and then on the beam current were symmetrically swapped [25]. In SPIDER, with $I_{PG}$ directed downwards, the plasma at the PG was observed to drift upwards [31]; the interplay between the magnetic filter field and the multicusp field at the source walls was however shown to affect the plasma properties in the eight independent RF drivers, in a way which is not symmetrical for an inversion of the magnetic filter field direction [44]. Further asymmetries on plasma and beam properties, in vertical and horizontal direction, are under study [45].

The negative ions are extracted from the source volume at the PG through 1280 apertures, whose position is represented in Figure 1b. During the experimental campaign whose data are here reported, most of the PG apertures were closed by a molybdenum mask, leaving only 80 apertures active; they are indicated in Figure 1b with red full circles. The purpose was to limit the gas outflow from the source, so as to keep the $H_2/D_2$ pressure around the source low enough to avoid breakdowns in the RF components [[4],[46]].

The extraction of negative ions is achieved by the electric field in the gap between the PG and the following grid, the extraction grid (EG); the electric field can penetrate through the PG apertures, draining negative ions. To counteract the full acceleration of electrons, magnets embedded in the EG produce alternated vertical magnetic fields that dump the co-extracted electrons onto the EG surface [1]; for the same purpose, the potential difference in the PG-EG gap, the extraction voltage ($U_{ex}$), is maintained by the ISEG (Ion Source – Extraction Grid) power supply at relatively low values (maximum ISEG ratings: 12 kV, 140 A [21]). The deflection of the negative ion trajectories caused by the co-extracted electron suppression (CESM) magnets in the EG is compensated by a further set of magnets in the GG [[47],[48]]. The acceleration of negative ions is completed in the gap between the EG and the Acceleration Grid (AG), which is kept at ground potential. The EG-AG potential difference, i.e. the acceleration voltage ($U_{acc}$), is set by the AGPS power supply (Acceleration Grid Power Supply, maximum ratings: 96 kV, 71 A [49]). Negative ions can then be fully accelerated up to 108 keV. In the experimental cases shown here, the limited beam current which is available in Cs-free conditions required to scale down $U_{ex}$ and $U_{acc}$, in order to maintain proper beam optics and beam transmission through the acceleration system [50].

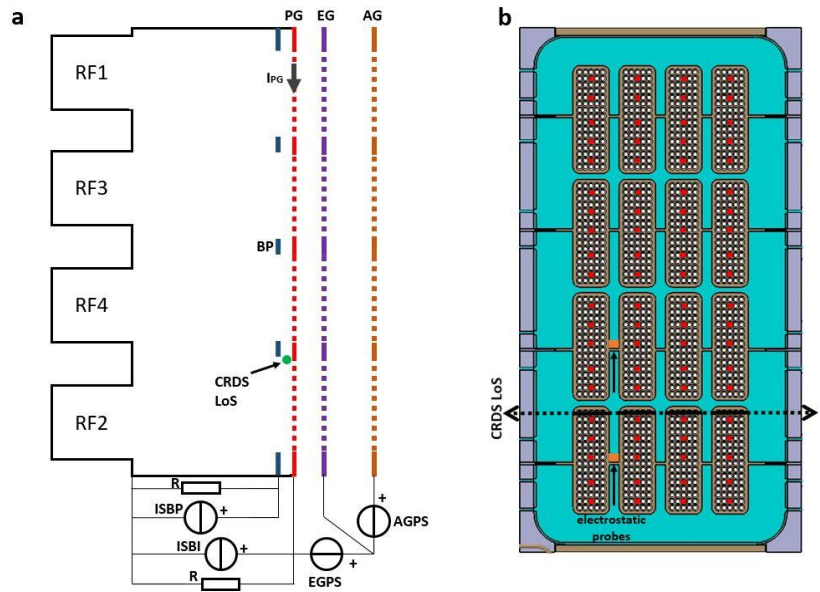

**Figure 1.** a) Scheme (vertical section) of the source and of the acceleration system of SPIDER, together with the electrical scheme of the main power supplies. b) Schematic representation of the BP (light blue) and of the PG, as seen from the back of the source; the active PG apertures are indicated with red dots, while the line of sight of the CRDS diagnostic is indicated in black. The two RF compensated electrostatic probes considered for the characterization of the plasma properties are indicated in brown; they are mounted on the rear side of the BP.

## 3. The CRDS diagnostic in SPIDER

The CRDS diagnostic in SPIDER [11] employs a Nd:YAG laser emitting beam pulses at 1064 nm wavelength, 6 ns duration and 150 mJ energy, at 10 Hz rate. The infrared laser is co-aligned to the 532 nm, 0.9 mW continuous beam of a laser diode, to facilitate the alignment of the optical setup. Both lasers are located in a room outside the SPIDER bioshield to protect them from neutron radiation. The co-aligned laser beams enter through an aperture in the concrete bioshield to reach the SPIDER vacuum vessel. Four mirrors steer the laser beams and direct them into the optical cavity. This is composed by two high reflectivity (>99.994%) mirrors, which act as vacuum-air interface, at two opposite flanges on the SPIDER vacuum vessel. Two vacuum tight structures, shown in Figure 2, provide the tilting and translation degrees of freedom required to align the two mirrors with respect to each other; the mirrors must also be aligned with respect to the 10 mm diameter apertures available on the source walls, so as to let the laser pulse pass from side to side through the centre of the holes. The maximum precision attainable in translations and tiltings is about 10 μm and 0.2 mrad, respectively. The two structures are also equipped with a gate valve to allow the substitution of the cavity mirrors while preserving the vacuum environment of SPIDER. Once trapped inside the optical cavity, the laser pulse crosses the plasma region back and forth several thousands times, undergoing partial absorption because of photodetachment reactions with negative ions, which adds up to the intrinsic losses of the cavity.

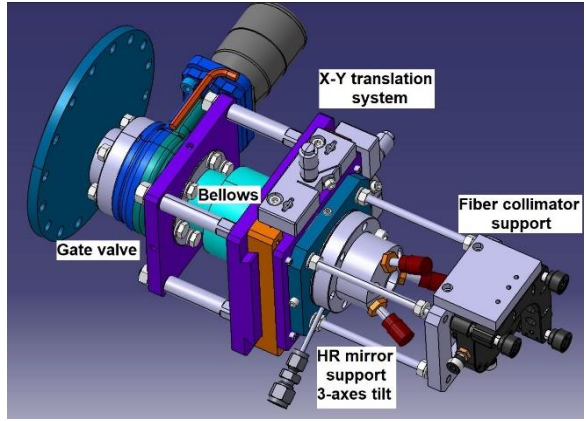

**Figure 2.** CAD model of the high reflectivity mirror holder, installed at the reception side.

The line of sight of the CRDS diagnostic, determined by the optical cavity, is located 5 mm from the PG upstream surface and it is horizontally oriented (Figure 1). The length of the optical cavity is L=4.637 m, while the region in which negative ions are present is estimated to be d=0.612 m long [11]. The position of the Line-of-Sight (LoS) with respect to the PG apertures is schematized in Figure 1b; the spacing between apertures is 20 mm horizontally and 22 mm vertically [1]. After the reduction of the number of active PG apertures, the LoS is not directly in front of active apertures, but 22 mm above four of them. Each time the laser beam in the cavity reaches the opposite mirror with respect to the injection side, a small fraction of it exits the cavity and is collected by a collimator, coupled to a 1 mm core silica fiber. The train of beam pulses exiting the cavity is then routed to the optical room, through a 1064 nm interference filter (10 nm FWHM bandpass) to block the spurious plasma light and finally to an avalanche photodiode (APD) detector. This has a 70 A/W response at 1064 nm, a $10^2$ internal gain and is coupled to a $4 \cdot 10^3$ V/A transimpedance amplifier. The detector output is acquired by a 250 MS/s digitizer. Figure 3 shows an example signal, collected for a CRDS laser shot in absence of plasma; the signal is reversed in polarity due to electronics. The light intensity decays exponentially, with a decay time $\tau$ estimated by a fitting function of the type $y = b - A \cdot \exp(-t/\tau)$, where b is the background level, A is the signal amplitude and t is time. The first 20 µs are normally not considered in the fit because affected by multiple excited cavity modes, leaving just TEM$_{00}$ (the mode with the smallest transverse dimensions) afterwards. In the example of Figure 3, $\tau$ is about 173 µs, which corresponds to about 5600 round trips of the laser pulse in the cavity; the CRDS signal can be considered extinguished in a time of about $5\tau$. The density of negative ions is estimated by comparing the decay time during the plasma phase, $\tau$, with the same quantity measured during plasma-off phases, $\tau_0$:

$$n_{H-} = \frac{L}{\sigma c d}\left(\frac{1}{\tau} - \frac{1}{\tau_0}\right)$$

where $c$ is the speed of light and $\sigma = 3.5 \cdot 10^{-21}$ m$^2$ is the photodetachment cross section at 1064 nm photon wavelength [51]. The typical measurement error is estimated to be about 10%. The analysis algorithm was optimized in order to provide negative ion density measurements in real time, at 10 Hz according to the laser pulse rate. The fluctuations of $\tau$-$\tau_0$ determine the minimum detection threshold of the diagnostic, which is presently around $2 \cdot 10^{15}$ m$^{-3}$; this value can be further reduced by averaging the measurements of density over significant time intervals. Proper techniques [11] are adopted to compensate for slow drifts of $\tau$ and $\tau_0$ with time, which are often observed during experimental days [52]. These drifts lead to a progressive reduction of the decay time, reducing the measurement range and increasing the statistical error of the measurements;

the original τ₀ value of the cavity is fortunately recovered during the night. The alignment stability of the laser head was proven to be able to provide non monothonic drifts of τ₀ [52]; the degradation of the decay time may be due to thermal expansion effects on the vacuum vessel and then on the mechanical supports of the CRDS optics. Similar phenomena of τ₀ degradation have been reported and attributed to a reduction of the cavity mirrors reflectivity, due to the deposition of impurities on their surfaces and to the UV radiation generated by the plasma [53]. The issue is not critical for the experimental conditions considered in this paper, i.e. plasma phases with duration of few tens of seconds. Nevertheless, it is under investigation to provide that CRDS will be able to deliver accurate real time measurements when plasma phases in SPIDER will reach a duration of hundreds or thousands of seconds.

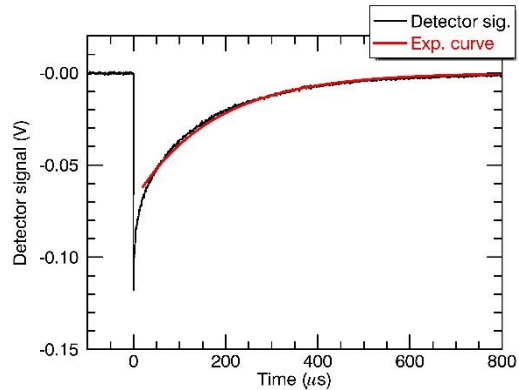

**Figure 3.** An example of CRDS signal acquired during a plasma-off phase, as function of time. The fitted exponential curve is shown in red. In this case, $\tau=173$ μs.

## 4. Experimental results

The CRDS diagnostic is the key tool to study how the negative ion density varies during the plasma phase, in particular as a consequence of the modification of source conditions or source/accelerator control parameters. The experimental results that follow were obtained in Cs-free conditions, the generation of negative ions and their space and velocity distributions are a product of the conditions of rovibrationally excited molecules flowing from the plasma bulk. The fluxes of Cs could then differ with respect to the future operation with Cs, in which most negative ions will be emitted from the surface as the result of the conversion of neutrals and ions, flowing from the plasma bulk toward the PG.

Sec. 4.1 presents the impact on the negative ion density of negative ion extraction from the PG apertures, within the present experimental conditions. Sec. 4.2 discusses the influence of the magnetic filter field and of gas pressure on negative ion density. Sec. 4.3 also discusses the influence of the magnetic filter field on the global beam performances; the comparison is further extended in sec. 4.4, comparing hydrogen and deuterium operation. At last, sec. 4.5 studies the effects of BP and PG biasing on the negative ion density and on the global beam performances, both in hydrogen and in deuterium.

### 4.1 Effects of beam extraction

According to the results of RF ion sources similar to SPIDER, with Cs evaporation the density of negative ions in proximity of the PG apertures decreases when a potential difference ($U_{ex}$) is applied between PG and EG, i.e. negatively charged particles are extracted from the ion source. [[13],[14]]. In the Cs-free experimental campaign under consideration, in most of the explored experimental conditions the density of H$^-$/D$^-$ was not affected by beam extraction. In some cases, however, the negative ion density resulted not to remain stable or get lower, but to increase during beam extraction; Figure 4 shows an example, the H$^-$/D$^-$ measurements being referred to a plasma pulse in deuterium, with source pressure p$_{source}$=0.3 Pa (as required for ITER HNB sources [2]), PG current I$_{PG}$=1.8 kA (2.9 mT magnetic filter field) and effective source body and BP bias currents I$_{BI}$=62 A and I$_{BP}$=43 A, respectively. The setting of the RF power was, for each generator: P$_{RF1}$=100 kW, P$_{RF2}$=100 kW, P$_{RF3}$=0 kW and P$_{RF4}$=100 kW; RF3 was off because of technical issues. The red shaded area in Figure 4 indicates the beam extraction phase (U$_{ex}$=1.83 kV, U$_{acc}$=22 kV). The phenomenon is surprising since the CRDS LoS is not in correspondence of a row of extraction apertures, but it is 22 mm above four active apertures.

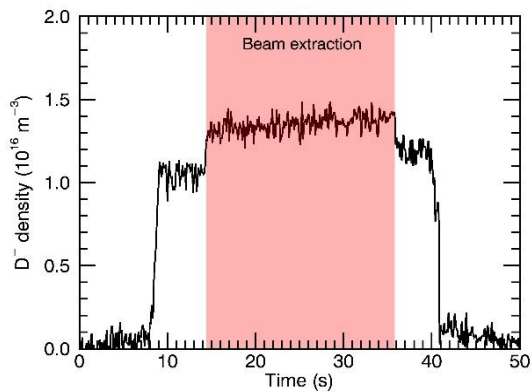

**Figure 4.** D$^-$ density as a function of time, in a plasma pulse with P$_{RF1}$=100 kW, P$_{RF2}$=100 kW, P$_{RF3}$=0 kW, P$_{RF4}$=100 kW, p$_{source}$=0.3 Pa, I$_{PG}$=1.8 kA, I$_{BI}$=62 A, I$_{BP}$=43 A. The beam extraction phase (U$_{ex}$=1.83 kV, U$_{acc}$=22 kV) is indicated by the red shaded area.

It was not possible to identify a clear and reproducible set of source parameters to exactly predict the appearance and impact of this phenomenon on negative ion density. Statistically, the increase of negative ion density with beam extraction was observed more frequently and with higher increase of negative ion density (up to about 50 %) in deuterium rather than in hydrogen. In hydrogen operation, this phenomenon usually appeared under modest RF power (≤240 kW total RF power) and low I$_{BI}$ and I$_{BP}$, usually leading the PG and the BP to have a negative bias current, and therefore to be below the plasma floating potential; in these cases, the amount of co-extracted electrons is higher.

A possible explanatory hypothesis is that the application of the extraction potential in the PG-EG gap leads to an increase of the potential in the plasma in front and around the PG active apertures, attracting more negative ions in that region; this would be possible because of the very limited number of active apertures, so that the local increase of the potential drains negative ions from the nearby plasma volume. At high bias, the effect would be less evident because the PG potential would already be above the plasma floating potential; moreover, the effect would be more evident with relatively low plasma density, for example because of low RF power. In a complementary way to this effect, the extraction of electrons from the active PG apertures may lead to an increase of the space charge, which allows some more negative ions to reach those same volumes; this would become much more evident when the ratio of electron to negative ion density is larger, i.e. in deuterium.

## 4.2 Influence of magnetic filter field and gas pressure

The Cs-free experimental campaign allowed to study the influence of the magnetic filter field intensity on the density of negative ions at the PG. Figure 5a shows the H$^-$ density, as measured by CRDS, as a function of the magnetic filter field intensity at the PG, with no beam extraction and biasing of BP and PG, at 400 kW RF power, uniformly distributed among the four RF generators, and three values of source filling pressure: 0.34 Pa, 0.39 Pa and 0.51 Pa. In these cases, the $I_{PG}$ direction was reversed, from downwards to upwards, with a consequent reversal of the magnetic filter field direction. As shown by the plot, the negative ion density at 0.34 Pa and 0.39 Pa source pressure is maximized at about 1.2÷1.6 mT filter field intensity, corresponding to an $I_{PG}$ range from about 0.7 kA to 1 kA; at 0.51 Pa the peaking of negative ion density is shifted at slightly higher values of magnetic filter field. The absolute values of negative ion density clearly increase with pressure.

A basic understanding of the relation between negative ion density and magnetic filter field intensity can be provided by considering the basic plasma properties at the PG. Figure 5b and Figure 5c show the values of plasma density and electron temperature as a function of the magnetic filter field intensity, for the same plasma pulse of Figure 5a. The reported values are averages from the measurements of two RF compensated electrostatic probes, mounted on the BP surface facing the plasma; their position, indicated in Figure 1b with brown rectangles, is above and below the CRDS LoS. The measurement errors are about 20% for plasma densities and 10 % for electron temperatures. The plasma density is found to be maximum at about 1.2 mT. A qualitative explanation of this behaviour is that the magnetic filter field, increasing from 0 mT, confines electrons (especially the faster ones) more and more towards the plasma drivers, as confirmed by the parallel reduction of electron temperature at the PG. As consequence, the plasma in the driver/rear region can produce and release a higher amount of ions, H/D neutrals and rovibrationally excited molecules towards the PG. If the magnetic filter field is too high, instead, it can penetrate deeper in the driver region [22], depressing the production of ions, neutrals and excited molecules.

The dependence of negative ion density on the magnetic filter field intensity results to be in qualitative agreement with the dependence of plasma density on the same parameter. This behavior can be explained by the fact that, in Cs-free conditions, the negative ions in the PG region are produced by dissociative attachment reactions which involve electrons and rovibrationally excited gas molecules (in turn produced, in the driver/rear expansion chamber, by electrons and gas molecules) [5]. Moreover, if the magnetic filter field intensity is below about 1 mT, the action of electron temperature reduction of the filter field is insufficient, increasing the rate of electron stripping reactions and then negative on losses. At last, measurements from the electrostatic probes show that the values of plasma density and electron temperature at the bias plate are rather insensitive to pressure variations in the range 0.34 Pa÷0.51 Pa. In contrast, the increase of the negative ion density with pressure is clear. This can be explained by the fact that the production rate of rovibrationally excited molecules, which then generate H$^-$/D$^-$ by dissociative attachment, grows with gas density. Moreover, it must be considered that, in absence of any PG biasing, the profile of plasma potential decreases towards the plasma electrode, so that negative charges are attracted backwards[32]. Increasing the pressure and then the gas density can increase the rate of collisions between negative ions and neutrals, redistributing the trajectories of H-/D- more favourably towards the PG. A more detailed investigation of negative ion production, destruction and transport in the SPIDER geometry will be made in future.

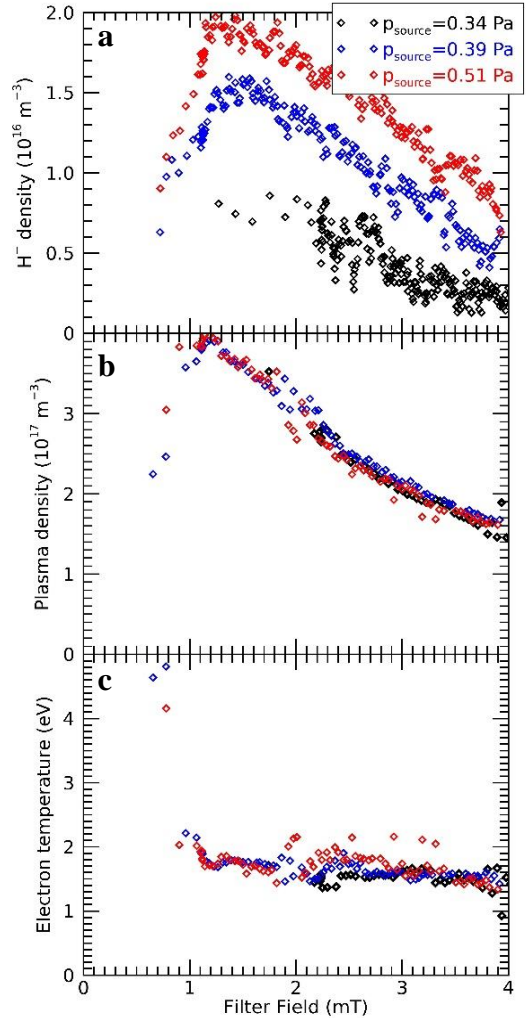

**Figure 5.** CRDS H- density measurements (plot a), plasma density (plot b) and electron temperature (plot c) measurements from the electrostatic probes as a function of the magnetic filter field intensity, under several operation conditions: all RF generators at 100 kW and source pressure at 0.34 Pa, 0.39 Pa and 0.51 Pa. No biasing of source and BP, no beam extraction. The magnetic filter field direction was reversed with respect to the standard configuration.

### 4.3 Magnetic filter field and beam performance

As explained in sec. 1, the experimentation on SPIDER should identify the operative conditions or practices that lead to maximize the ion beam current and minimize the fraction of co-extracted electrons. Besides limiting H-/D- losses by electron stripping, the magnetic filter field has the purpose of reducing the amount of co-extracted electrons, as demonstrated in other experiments [35]; its influence on the negative ion density, shown in sec. 4.2, also affects the beam current. In Figure 6 the negative ion density, the electron current density $j_e$, the extracted beam current density $j_H$ and the fraction of co-extracted electrons $j_e/j_H$ are plotted as functions of the magnetic filter field intensity at the PG, for hydrogen plasma operation; $I_{PG}$ was flowing in the standard direction, i.e. downwards. With beam extraction, in Cs-free operation the filter field had to necessarily exceed 1.6 mT in order to limit the amount (and consequently the total power) of co-extracted electrons collected by the EG, in turn to prevent heat load damages on the grid itself. The operating conditions were: $P_{RF1}$=100 kW, $P_{RF2}$=100 kW, $P_{RF3}$=0 kW (technical issues on RF3), $P_{RF4}$=100 kW, $p_{source}$=0.33 Pa, no biasing of source and BP, $U_{ex}$=1.5 kV, $U_{acc}$=20 kV. In the present conditions the electron current density $j_e$ was estimated as the ratio between the EGPS current and the total area of the active PG apertures; the EGPS current also includes the AGPS current, but in Cs-free conditions it is negligible compared to the current flowing in the EG. To provide a conservative estimation of the H$^-$/D$^-$ current density $j_H - j_D$, the APGS current was multiplied by a factor 0.4; this correction coefficient was estimated by a comparison of beam electric and calorimetric measurements, to measure the effective amount of negative ions in the beam without including the contribution of secondary electrons [54]. The applied extraction voltage $U_{ex}$=1.5 kV was such that the negative ion beam current is already at saturation, i.e. it was not limited by the space charge in the accelerator according to the Child Langmuir law, but by the flux of negative charged particles which are available at the meniscus of each PG aperture.

In relative terms, comparing Figure 5a and Figure 6a it results that the dependence of the negative ion density on $I_{PG}$ is the same with and without extraction and independently from the magnetic filter field direction. From 1.6 mT onwards, the negative ion density decreases with the magnetic filter field intensity, and so does the extracted beam current density. Figure 7, for the same experimental data, shows that the extracted current density has an essentially linear dependence on negative ion density, similarly to what found elsewhere with Cs evaporation [13].

Figure 6b shows that, as expected, the co-extracted electron current density decreases with the magnetic filter field. However, $j_e/j_H$ is not minimized at the same condition that maximizes the negative ion density availabilty, but in a broad range of magnetic field intensities centered at 3.2 mT ($I_{PG}$~2 kA). A filter field strength between 1.4 mT and 3.2 mT, respectively providing the maximum negative ion density and extracted current and a minimum fraction of co-extracted electrons, can be considered as a valid starting point for the subsequent experimentation with Cs evaporation. Even though the production of negative ions will be dominated by different processes, the depth of the cold-electron region before the PG should depend on the filter field in a similar fashion, thus realizing an optimal condition for negative ion survival in the same range of $I_{PG}$. The reduction of plasma density at high filter field (and then the availability neutrals and positive ions) is still expected to happen with Cs evaporation; and, on the other hand, the co-extracted electron current [5] will largely decrease thanks to the higher density of negative ions at the PG surface, that will significantly hinder the transport of electrons. For these reasons, the filter field strength required to maximise the plasma density and minimise perpendicular drifts is expected to be similar or lower than what found in Cs-free conditions, at least in hydrogen operation.

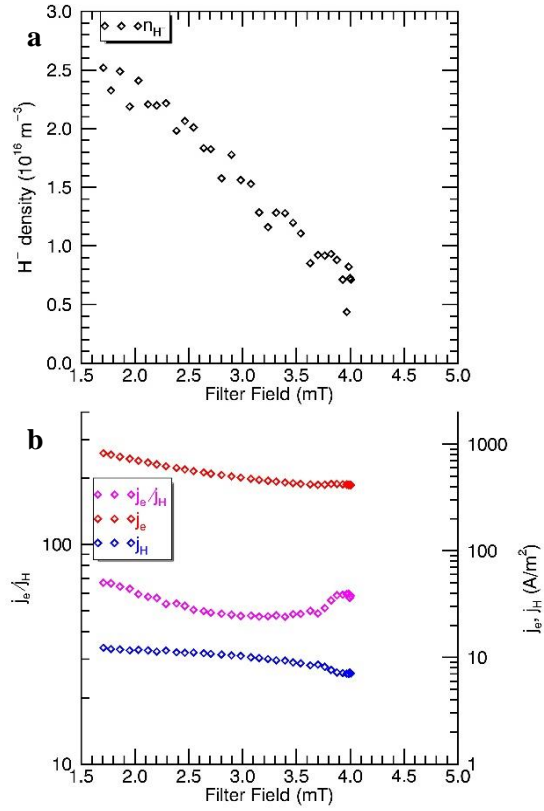

**Figure 6.** Negative ion density (black), co-extracted electron current density $j_e$ (red), extracted beam current density $j_H$ (blue) and ratio of co-extracted electrons $j_e/j_H$ (purple) as a function of the magnetic filter field intensity at the PG. Operation conditions: $P_{RF1}$=100 kW, $P_{RF2}$=100 kW, $P_{RF3}$=0 kW, $P_{RF4}$=100 kW, $p_{source}$=0.33 Pa (hydrogen), no biasing of source and BP, $U_{ex}$=1.5 kV, $U_{acc}$=20 kV. $I_{PG}$ was in standard direction.

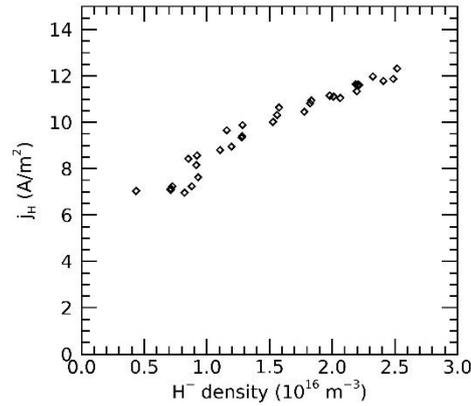

**Figure 7.** Extracted beam current density $j_H$ as a function of the negative ion density. The data and the experimental conditions are the same as in Figure 8.

## 4.4 Isotopic effect on the magnetic filter field action

The dependence of negative ion density and beam performances on the magnetic filter field intensity must also be compared between hydrogen and deuterium plasmas. SPIDER and the future ITER HNBs are indeed required to operate with both gases; isotopic effects on the source plasma and on the beam extraction have been object of several studies, as described by Bacal and Wada [55]. Figure 8 shows the negative ion density, the electron current density $j_e$, the extracted beam current density $j_D$ and the fraction of co-extracted electrons $j_e/j_D$ as functions of the magnetic filter field intensity, in a plasma pulse in deuterium with the same operative conditions of the data in Figure 6 (the pressure was slightly lower, 0.29 Pa).

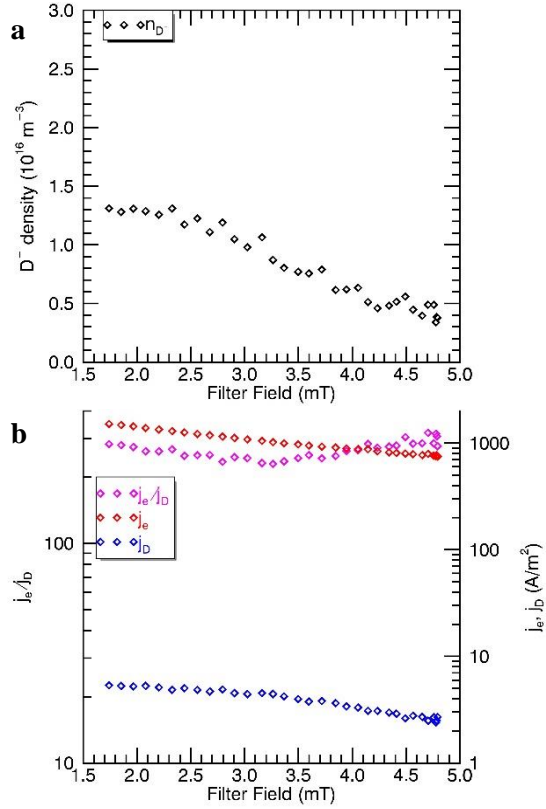

**Figure 8.** Negative ion density (black), co-extracted electron current density $j_e$ (red), extracted beam current density $j_D$ (blue) and ratio of co-extracted electrons $j_e/j_D$ (purple) as a function of the magnetic filter field intensity at the PG, for a plasma pulse in deuterium. The SPIDER operation conditions are the same as in Figure 6, $p_{source}$ was slightly lower (0.29 Pa).

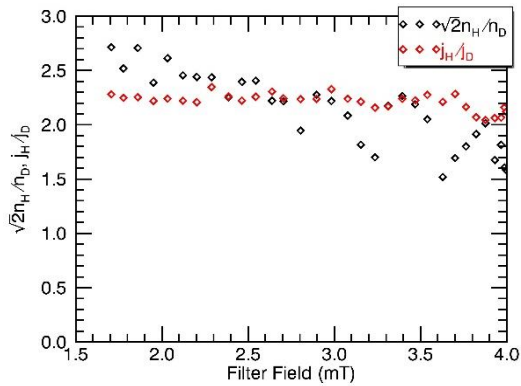

**Figure 9.** Ratio between the negative ion density measurements in hydrogen and deuterium (corrected by a factor √2), and ratio between the extracted current density in hydrogen and deuterium, as a function of the magnetic filter field intensity. The data belong to the cases shown in Figure 6 and Figure 8; the data in deuterium have been interpolated over the same $I_{PG}$ values of the data in hydrogen.

In relative terms the value of magnetic filter field intensity, which is required to maximize negative ion density and extracted beam current density, is the same in hydrogen and in deuterium. Similarly, the fraction of co-extracted electrons is minimized in deuterium in the same magnetic field intensity range that was required in hydrogen. Comparing Figure 6 and Figure 8, it results that around 1.6 mT the H⁻ density is almost twice the D⁻ density available at the same conditions. This difference however reduces as the filter field increases. The slight difference in source pressure between the hydrogen and deuterium cases (-13 % in D) cannot explain the such relevant reduction of negative ions in deuterium. The phenomenon was also observed in previous experiments, as described by Bacal and Wada [55]. The cause was hypothesized to be a combination in deuterium of lower density of rovibrationally excited molecules (then lower production of negative ions), together with a higher availability of neutrals at the PG, which would lead to a higher associative detachment reaction (D⁻+D→$D_2$+e) rate and then more negative ion losses. These hypothese will be object of future verification in SPIDER through the combined use of source diagnostics.

Regarding the currents extracted through the grids at the same extraction voltage, in deuterium $j_e$ is higher and $j_D$ is lower than the corresponding values in hydrogen, so that $j_e/j_D$ is around $10^3$, while $j_e/j_H$ is around few tens. This was observed also in other similar sources [[35],[39],[56]].

A basic relation between the ion density and current measurements in hydrogen and deuterium can be established. The characterization of the beam optics in both hydrogen and deuterium indicated that at $U_{ex}$=1.5 kV, as in the cases of Figure 6 and Figure 8, in both gases the beam current is only constrained by the negative ion flux towards the meniscus of the PG apertures. The beam current density can then be assumed to be proportional to the negative ion density and to the average negative ion velocity; the latter shall be proportional to $\sqrt{T/m}$, with T a temperature, and m the mass of negative ions. Assuming the same temperature rule the flow of H⁻ and D⁻, it is easily derived that $j_H/j_D$ should be comparable to √2·$n_H/n_D$. This is verified in Figure 9, showing these two quantities as function of the magnetic filter field intensity. The data belong to the same two SPIDER shots considered in Figure 8; to compute the ratios, $n_D$ and $j_D$ where interpolated over the magnetic filter field intensity values of the corresponding measurements in hydrogen. $j_H/j_D$ results to be compatible with √2·$n_H/n_D$ ,confirming the validity of this scaling in the Cs-free sources and in the experimental conditions considered here. The agreement between $j_H/j_D$ and √2·$n_H/n_D$ also confirms that the temperature T mentioned before is roughly the same in the magnetic filter field scans in hydrogen and deuterium.

### 4.5 Effects of BP and PG biasing

The effect of bias currents on H⁻/D⁻ density was also studied; at this early stage of investigation, the cumulative effect of $I_{BI}$ and $I_{BP}$ was considered. Figure 10 shows the H⁻ density and the D⁻ density, together with the respective values of $j_H - j_D$, $j_e$ and $j_e/j_H - j_e/j_D$, as a function of $I_{BI}+I_{BP}$. In both cases $P_{RF1}$=100 kW, $P_{RF2}$=100 kW, $P_{RF3}$=0 kW, $P_{RF4}$=100 kW, $p_{source}$=0.3 Pa, $I_{PG}$=2.0 kA downwards (corresponding to 3.2 mT at the PG), $U_{ex}$=2.5 kV and $U_{acc}$=25 kV. In the case of deuterium, the technical conditions of the source and the limited time of the deuterium experimental campaign constrained $I_{BI}+I_{BP}$ to a range in which the bias current direction is mostly reversed. As observed in sec. 4.4, in hydrogen the extracted current density is higher, while $j_e$ and $j_e/j_H$ are generally lower than in deuterium. In the case of hydrogen, the increase of bias currents is steadily beneficial in reducing $j_e/j_H$, mainly because of the lowering of $j_e$, while the extracted current density grows with the bias current in the negative range and saturates in the positive one, with a slight peak between 0 A and 100 A. Similarly, the H$^-$ density grows with $I_{BI}+I_{BP}$ in the negative range and is slightly peaked in between 0 A and 100 A. In the case of deuterium, the effects of $I_{BI}+I_{BP}$ on negative ion density, $j_D$, $j_e$ and $j_e/j_D$ are qualitatively similar to what observed in hydrogen in the same current range.

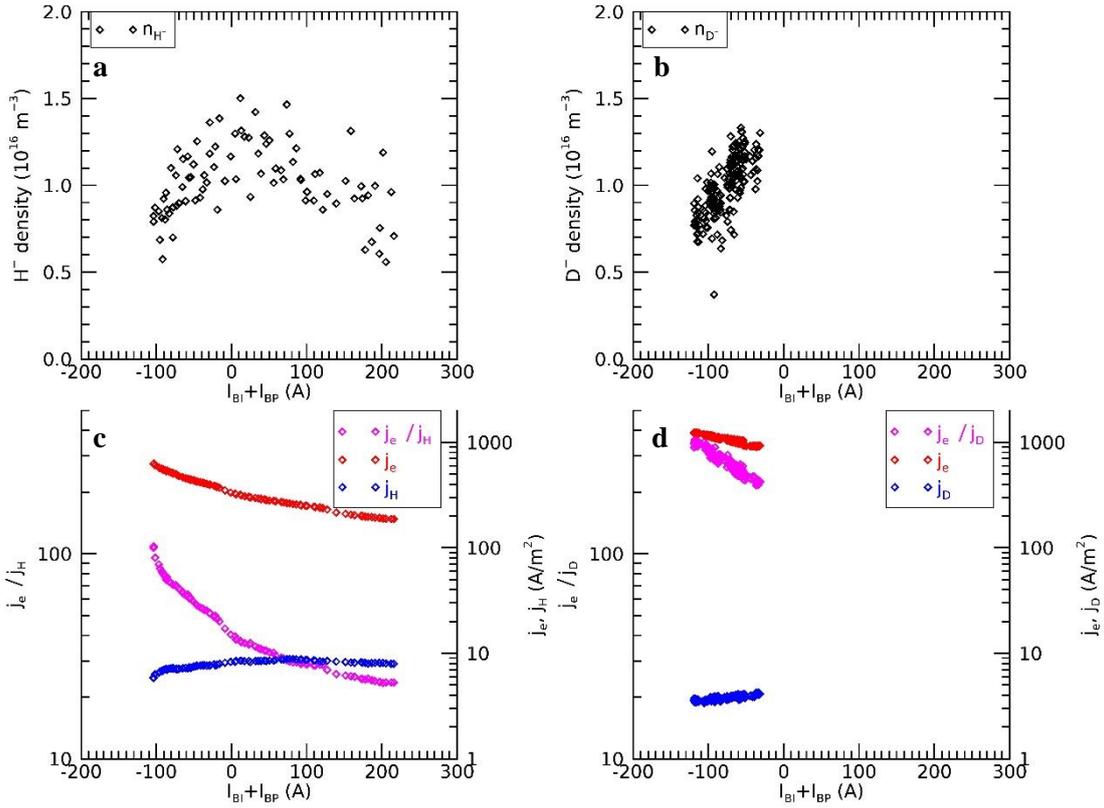

**Figure 10.** Negative ion density (black), co-extracted electron current $I_e$ (red), extracted beam current $I_b$ (blue) and ratio of co-extracted electrons $j_e/j_H$ (purple) as a function of the sum of bias currents $I_{BI}$ and $I_{BP}$. Plot a and c show these physical quantities for a plasma pulse in hydrogen, while plots b and d show analogous measurements for a deuterium plasma pulse. Common operation conditions for both cases: $P_{RF1}$=100 kW, $P_{RF2}$=100 kW, $P_{RF3}$=0 kW, $P_{RF4}$=100 kW, $p_{source}$=0.3 Pa, $I_{PG}$=2.0 kA (current in standard direction, equivalent to 3.2 mT), $U_{ex}$=2.5 kV, $U_{acc}$=25 kV.

## 5. Conclusions

The negative ion sources of SPIDER and of the future ITER HNBs are required to reach ambitious performances in terms of maximizing the beam current density and minimizing the fraction of co-extracted electrons. Beam current is strongly related to the negative ion density at the PG apertures, quantity that can be monitored in SPIDER by the CRDS diagnostic. Most of the source parameters that can affect the negative ion density have been discussed in this paper, showing which Cs-free operating conditions can provide the highest negative ion density and beam current. These results will constitute the basis for the further optimization of SPIDER with Cs evaporation, to meet ITER requirements.

This paper confirms the importance of gas pressure in determining the negative ion density, which is however expected to be fixed in the multi-grid accelerator of the ITER HNB. CRDS data also showed that in SPIDER, between about 0.3 Pa and 0.5 Pa source pressure, an increase of the magnetic filter field intensity would not lead to a monotonic improvement of performances; an optimum negative ion density is found at about 1.4 mT÷1.6 mT (at the PG). This effect is due to the peaking of plasma density in the extraction region at the same value of filter field; both phenomena are due to the confinement of fast electrons in the driver/rear region because of the magnetic field. Moreover, by lowering the filter field intensity the reduction of plasma density is also accompanied by an increase of the electron temperature, further limiting the negative ion survival probability in the plasma volume close to the PG.

The extracted beam current density resulted to vary linearly with the negative ion density for magnetic filter field variations in the range 1.6 mT÷4.0 mT. As a result, the beam current density is maximum at about 1.6 mT, both in hydrogen and deuterium. A compromise value however must be adopted as a balance with the requirement of limiting the fraction of co-extracted electrons, which is best satisfied at about 3.2 mT, in both hydrogen and deuterium. Intermediate values between the two settings can still provide adequate performances.

Regarding bias currents, their effect on the fraction of co-extracted electrons was always beneficial, while H$^-$ ion density is maximized at $I_{BI}+I_{BP}$ values between 0 A and 100 A. At last, it was shown that in certain experimental conditions the beam extraction can lead to an increase of negative ion density at the PG apertures, phenomenon which may help understanding the supply of volume-generated negative ions in the proximity of the PG. For a deeper interpretation of CRDS measurements in Cs-free operation, it will be necessary to globally consider the information from all the source diagnostic, and simulate the plasma reactions and transport taking into account the entire source geometry.

**Acknowledgments**

This work has been carried out within the framework of the ITER-RFX Neutral Beam Testing Facility (NBTF) Agreement and has received funding from the ITER Organization. The views and opinions expressed herein do not necessarily reflect those of the ITER Organization.

This work has been carried out within the framework of the EUROfusion Consortium and has received funding from the Euratom research and training programme 2014-2018 and 2019-2020 under grant agreement No 633053. The views and opinions expressed herein do not necessarily reflect those of the European Commission.
This work was supported in part by the Swiss National Science Foundation.